\documentclass{article}
\usepackage{graphicx}
\usepackage{amsmath}
\usepackage{amsfonts}
\usepackage{amssymb}

\begin{document}

\author{Antony Valentini\\Augustus College}

\begin{center}
{\LARGE Hidden Variables, Statistical Mechanics and the Early Universe}

\bigskip Antony Valentini$^{0a}$\footnotetext{$^{a}$\textit{Present address}:
Theoretical Physics Group, Blackett Laboratory, Imperial College, London SW7
2BZ, England, and Center for Gravitational Physics and Geometry, Department of
Physics, The Pennsylvania State University, University Park, PA 16802, USA.
\textit{Permanent address}: Augustus College, 14 Augustus Road, London SW19
6LN, England.}
\end{center}

\section{Introduction}

One of the central mysteries of quantum theory is that it seems to be
fundamentally nonlocal -- and yet the nonlocality cannot be used for practical
signalling at a distance. As argued elsewhere (Valentini 1991a,b; 1992; 1996),
the consistency of modern physics seems to depend on a `conspiracy', in which
nonlocality is hidden by quantum noise. It is as if there is an underlying
nonlocality which we are unable to control because of the statistical
character of quantum events.

A natural explanation for this peculiar state of affairs is the hypothesis
that quantum probability distributions are not fundamental, but merely
represent a special state of equilibrium in which nonlocality happens to be
masked by statistical noise (Valentini 1991a,b; 1992; 1996). In quantum theory
a system with wavefunction $\psi$ has an associated probability distribution
given by the Born rule $\rho=|\psi|^{2}$. This is usually regarded as a
fundamental law. If instead we regard $\rho=|\psi|^{2}$ as analogous to, say,
the Maxwell distribution of molecular speeds for a classical gas in thermal
equilibrium, the above `conspiracy' takes on a different light: it is seen to
be an accidental, contingent feature of the `quantum equilibrium' distribution
$\rho=|\psi|^{2}$. In a universe that is in (global) thermal equilibrium, it
is impossible to convert heat into work; this is the classical thermodynamic
heat death. In a universe that is everywhere in quantum equilibrium, it is
impossible to use nonlocality for signalling; this is the `subquantum heat
death' which, we claim, has actually occurred in our universe.

The pilot-wave theory of de Broglie and Bohm (de Broglie 1928; Bohm 1952a,b)
offers a concrete model of this scenario.\footnote{Note that in 1927 at the
Fifth Solvay Congress de Broglie proposed the full pilot-wave dynamics in
configuration space for a many-body system, not just the one-body theory; and
unlike Bohm's reformulation of 1952, de Broglie's new approach to dynamics in
the 1920s was always based on velocities (Valentini 2001a). The proceedings of
the Fifth Solvay Congress are being translated into English (Bacciagaluppi and
Valentini 2001).} A system with wavefunction $\psi$ is assumed to have a
definite configuration $x(t)$ at all times whose velocity is determined by the
de Broglie guidance equation $\pi=\partial S/\partial x$, where $\pi$ is the
canonical momentum and $S$ is the phase of $\psi$ (given locally by
$S=\hslash\operatorname{Im}\ln\psi$). The wavefunction $\psi$ is interpreted
as an objective `guiding field' in configuration space, and satisfies the
usual Schr\"{o}dinger equation.\footnote{Pilot-wave dynamics may be applied to
fields as well as particles -- even to fermion fields (Valentini 1992; 1996;
2001a) and non-Abelian gauge theories (Valentini 2001a). Note that the natural
spacetime structure associated with this velocity-based dynamics is
Aristotelian spacetime $E\times E^{3}$, which has an inbuilt natural state of
rest (Valentini 1997).}

At the fundamental hidden-variable level, pilot-wave theory is nonlocal. For
example, for two entangled particles A and B the wavefunction $\psi
(x_{A},x_{B},t)$ has a non-separable phase $S(x_{A},x_{B},t)$ and the velocity
$dx_{A}/dt=\nabla_{A}S(x_{A},x_{B},t)/m$ of particle A depends instantaneously
on the position $x_{B}$ of particle B.\footnote{It is being assumed here that
the position of particle B could in some sense be changed while keeping the
wavefunction fixed.} However at the $\mathit{quantum}$ level, where one
considers an ensemble with distribution $\rho(x_{A},x_{B},t)=|\psi(x_{A}%
,x_{B},t)|^{2}$, operations at B have no statistical effect at A: as is well
known, quantum entanglement cannot be used for signalling at a distance.

It is worth emphasising that this `washing out' of nonlocality by statistical
noise is peculiar to the distribution $\rho=|\psi|^{2}$. If one considers an
ensemble of entangled particles at $t=0$ with distribution $\rho_{0}%
(x_{A},x_{B})\neq|\psi_{0}(x_{A},x_{B})|^{2}$, it may be shown by explicit
calculation that changing the Hamiltonian at B induces an instantaneous change
in the marginal distribution $\rho_{A}(x_{A},t)\equiv%
{\textstyle\int}
dx_{B}\ \rho(x_{A},x_{B},t)$ at A. For a specific example it was found that a
sudden change $\hat{H}_{B}\rightarrow\hat{H}_{B}^{\prime}$ at B -- say a
change in potential -- leads after a short time $\epsilon$ to a change
$\Delta\rho_{A}\equiv\rho_{A}(x_{A},\epsilon)-\rho_{A}(x_{A},0)$ at A given by
(Valentini 1991b)
\begin{equation}
\Delta\rho_{A}=-\frac{\epsilon^{2}}{4m}\frac{\partial}{\partial x_{A}}\left(
a(x_{A})\int dx_{B}\ b(x_{B})\frac{\rho_{0}(x_{A},x_{B})-|\psi_{0}(x_{A}%
,x_{B})|^{2}}{|\psi_{0}(x_{A},x_{B})|^{2}}\right)
\end{equation}
(Here $a(x_{A})$ depends on $\psi_{0}$; $b(x_{B})$ also depends on $\hat
{H}_{B}^{\prime}$ and is zero if $\hat{H}_{B}^{\prime}=\hat{H}_{B}$.)

For nonequilibrium ensembles $\rho_{0}\neq|\psi_{0}|^{2}$, there are genuine
\textit{instantaneous signals} at the statistical level. This has recently
been shown to be a general feature, independent of pilot-wave theory. Any
deterministic hidden-variables theory that reproduces quantum theory leads,
for hypothetical nonequilibrium ensembles, to instantaneous signals at the
statistical level (Valentini 2001b).

Note that our inability to see the trajectories -- or hidden variables -- is
also a contingent feature of equilibrium: the uncertainty principle holds if
and only if $\rho=|\psi|^{2}$ (Valentini 1991b).\footnote{Note, however, that
the approximately classical functioning of the experimenter also plays a role
here. A subquantum demon -- an automaton that functions at the hidden-variable
level -- would see subquantum trajectories even for an equilibrium ensemble
(Valentini 2001a).} Heuristically, it is natural to compare our limitations
with those of a Maxwell demon in thermal equilibrium with a gas, whose
attempts to sort fast and slow molecules fail. The common objection to hidden
variables -- that their detailed behaviour can never be observed, making their
existence doubtful -- is seen to be misguided: for the theory cannot be blamed
if we happen to live in a state of statistical equilibrium that masks the
underlying details. There is no reason why nonequilibrium could not exist in
the remote past or in distant regions of the universe (Valentini 1992), in
which case the details of the `hidden-variable level' would not be hidden at all.

Our central theme, then, is the subquantum `heat death of the universe' at the
hidden-variable level. How does one account for the quantum noise --
encapsulated by the Born rule $\rho=|\psi|^{2}$ -- that pervades our observed
universe at the present time?

\section{\bigskip`Empirical' Approach to Statistical Mechanics}

It will be shown how one may set up an analogue of classical statistical
mechanics, based on pilot-wave dynamics in configuration space (Valentini
1991a; 1992; 2001a). We shall focus in particular on an analogue of the
classical coarse-graining approach. But first, we shall address some
foundational issues that were not considered in the author's original papers.

In \textit{what sense} might one `explain' the probability distribution
$\rho=|\psi|^{2}$, in the context of a deterministic (and time-reversal
invariant) pilot-wave theory?

Say the Moon is now at position $P$ (in phase space) at time $t$. This might
be `explained' -- using Newton's laws -- by the fact that the Moon was
(earlier) at position $Q$ at time $t_{0}$; and if $t_{0}$ is in the remote
past, one would in practice \textit{deduce} that the Moon must have been at
$Q$ at time $t_{0}$, from the observed position $P$ today. That one has had to
deduce the past ($Q$) from the present ($P$) would not change our physical
intuition that the Moon may reasonably be said to be at $P$ now `because' it
was at $Q$ at $t_{0}$.

This example may seem remote from statistical mechanics. However, many workers
would object -- in the author's opinion wrongly -- that $Q$ at $t_{0}$ is a
mere deduction, and \textit{cannot} be regarded as a satisfactory explanation
for $P$ today. The argument would be that one would have a satisfactory
explanation only if it could be shown that \textit{all} -- or at least `most'
-- possible positions of the Moon at $t_{0}$ evolve into the Moon being at $P$
today. Only then would one have an `explanation' for the present, as opposed
to a mere deduction about the past.

If this seems an unfair portrayal of what many workers in the foundations of
statistical mechanics would claim, then consider what is attempted in that
field. The `Holy Grail' has always been to explain relaxation to thermal
equilibrium by showing that all or `most' initial states do so. And the great
stumbling block has always been that, by time-reversal invariance, for every
initial state that evolves towards equilibrium there is another that evolves
away from it. One must then place restrictions on initial states, such as an
absence of correlations, or a lack of fine-grained microstructure, and so on,
restrictions that are violated by the unwanted `time-reversed',
entropy-decreasing states. One then tries to argue that the restricted set of
initial states is `natural', or that `most' initial states -- with respect to
some suitable measure -- satisfy the required conditions. As is well known,
this program is as controversial today as it was a century ago.

To avoid such controversy one might adopt a more modest -- and more realistic
-- `empirical' approach (Valentini 1996; 2001a). Going back to the example of
the Moon, one would try to deduce -- or perhaps guess -- where the Moon must
have been before to explain its position today. In the same sense, for a box
of gas evolving to thermal equilibrium, one should try to deduce or guess what
the initial (micro-)state must have been like to yield the observed behaviour
-- without trying (in vain) to show that all or `most' initial states would do
so. Of course, in the case of a gas there will be a whole class of microstates
yielding the required behaviour; one will not attempt to deduce the exact
initial state uniquely (unlike in the case of the Moon). Further, because
there are so many variables one resorts to statistical methods. Thus, one
tries to construct a class of initial conditions that yields the observed
behaviour, and one tries to understand the evolution of that class towards a
unique equilibrium state on the basis of a general mechanism (without having
to solve the exact equations of motion).

In the case at hand, what is to be explained is the observation of equilibrium
today to within a certain experimental accuracy, $\rho=|\psi|^{2}\pm\epsilon$.
For example, for a large number of independent Hydrogen atoms each with
ground-state wavefunction $\psi_{100}$, one might measure the distribution of
electron positions (with respect to the nuclei, in a nonrelativistic model) as
accurately as current technology allows, and obtain a result $\rho=|\psi
_{100}|^{2}\pm\epsilon$. The initial or earlier conditions of the universe
must have been such as to reproduce this result today.

One possible initial condition is just equilibrium itself (which is preserved
by the equations of motion). That is, the universe could have started in
equilibrium, leading to $\rho=|\psi|^{2}$ exactly today. But initial
equilibrium is only one possibility among (uncountably) many: even without any
proofs or assumptions about relaxation $\rho\rightarrow|\psi|^{2}$, it is
clear on grounds of continuity alone that if initial equilibrium evolves to
equilibrium today then an infinite class of initial states sufficiently close
to equilibrium must evolve to $\rho=|\psi|^{2}\pm\epsilon$ today; and given
the violence of the early universe, one expects there to be a large class of
initial states \textit{far} from equilibrium that evolve to $\rho=|\psi
|^{2}\pm\epsilon$ today as well.

What sort of past conditions could have evolved to $\rho=|\psi|^{2}\pm
\epsilon$ today? Below, we shall hypothesise a class of possible initial
(nonequilibrium) states and give a general mechanism explaining their approach
to equilibrium, as is done in classical statistical mechanics.

It is not true, of course, that all initial states evolve towards equilibrium;
time-reversal invariance forbids. Nor does it need to be true: `bad' initial
conditions are ruled out on \textit{empirical}, observational grounds.

\section{Subquantum \textit{H}-Theorem}

There are many approaches to classical statistical mechanics. Here we focus on
a pilot-wave analogue of the classical coarse-graining \textit{H}-theorem
(Valentini 1991a). This is not because we believe coarse-graining to be the
best approach; it has advantages and disadvantages, like the others.

For a classical isolated system, both the probability density $p$ and the
volume element $d\Omega$ (on \textit{phase} space) are preserved along
trajectories. The classical \textit{H}-function $H_{class}=\int d\Omega\ p\ln
p$ is constant in time. If we replace the fine-grained $p$ by the
coarse-grained $\bar{p}$ and assume that $\bar{p}_{0}=p_{0}$ at $t=0$, then
$\bar{H}_{class}(t)\leq\bar{H}_{class}(0)$ -- the classical coarse-graining
\textit{H}-theorem of the Ehrenfests. The decrease of $\bar{H}_{class}$
corresponds to the formation of structure in $p$ and the consequent approach
of $\bar{p}$ to uniformity. (See for example Tolman (1938).)

Consider now an ensemble of complicated many-body systems, each with
wavefunction $\Psi$, the configurations $X$ distributed with a probability
density $P$. The gradient $\nabla S$ of the phase of $\Psi$ determines a
velocity field $\dot{X}$ in configuration space. (For a low-energy system of
$N$ particles of mass $m$, we have 3-vector velocities $\dot{X}_{i}=\nabla
_{i}S/m$ where $i=1,2,....,N$.) The continuity equations
\begin{equation}
\frac{\partial P}{\partial t}+\nabla\cdot(\dot{X}P)=0
\end{equation}
(which holds by definition of $P$) and
\begin{equation}
\frac{\partial|\Psi|^{2}}{\partial t}+\nabla\cdot(\dot{X}|\Psi|^{2})=0
\end{equation}
(which follows from the Schr\"{o}dinger equation) imply that the ratio
$f\equiv P/|\Psi|^{2}$ is preserved along trajectories: $df/dt=0$, where
$d/dt=\partial/\partial t+\dot{X}\cdot\nabla$. Further, like $d\Omega$
classically, the element $|\Psi|^{2}dX$ is preserved along trajectories too
(where $dX$ is the volume element in configuration space). This suggests
replacing $p\rightarrow f$ and $d\Omega\rightarrow|\Psi|^{2}dX$ in $H_{class}%
$, yielding the \textit{subquantum H-function}
\begin{equation}
H=\int dX\ P\ln(P/|\Psi|^{2})
\end{equation}
This is in fact the relative negentropy of $P$ with respect to $|\Psi|^{2}$.

The above continuity equations now imply that $H=H(t)$ is constant: $dH/dt=0$,
as in the classical case. But if one divides configuration space into cells of
volume $\delta V$ and averages $P$ and $|\Psi|^{2}$ over the cells, yielding
coarse-grained values $\bar{P}$ and $\overline{|\Psi|^{2}}$, one may define a
coarse-grained \textit{H}-function
\begin{equation}
\bar{H}=\int dX\ \bar{P}\ln(\bar{P}/\overline{|\Psi|^{2}})
\end{equation}
If we assume the initial state has `no fine-grained microstructure' at $t=0$,%
\begin{equation}
\bar{P}_{0}=P_{0}\ ,\;\;\;\overline{|\Psi_{0}|^{2}}=|\Psi_{0}|^{2}%
\end{equation}
then it may be shown that%
\begin{equation}
\bar{H}_{0}-\bar{H}(t)=\int dX\ |\Psi|^{2}\left(  f\ln(f/\tilde{f})+\tilde
{f}-f\right)  \geq0
\end{equation}
(where $\tilde{f}\equiv\bar{P}/\overline{|\Psi|^{2}}$), so that
\begin{equation}
\bar{H}(t)\leq\bar{H}_{0}%
\end{equation}
for all $t$ (Valentini 1991a).

It is instructive to examine the time derivatives of $\bar{H}(t)$ at $t=0$.
Use of the continuity equations shows that $\left(  d\bar{H}/dt\right)
_{0}=0$ and (Valentini 1992)%
\begin{equation}
\left(  \frac{d^{2}\bar{H}}{dt^{2}}\right)  _{0}=-\int dX\frac{|\Psi_{0}|^{2}%
}{f_{0}}\left(  \overline{(\dot{X}_{0}\cdot\nabla f_{0})^{2}}-\overline
{(\dot{X}_{0}\cdot\nabla f_{0})}^{2}\right)  \leq0
\end{equation}
The quantity in brackets is just the (non-negative) variance var$_{\delta
V}(\dot{X}_{0}\cdot\nabla f_{0})$ of $(\dot{X}_{0}\cdot\nabla f_{0})$ over a
coarse-graining cell $\delta V$. If we suppose that var$_{\delta V}(\dot
{X}_{0}\cdot\nabla f_{0})\neq0$ then we have the strict inequality $\bar
{H}(t)<\bar{H}_{0}$ and $\bar{H}(t)$ \textit{must decrease} over at least a
finite time interval $(0,T)$ immediately after $t=0$ (Valentini 2001a).

Because $x\ln(x/y)+y-x\geq0$ for all real $x,y$, the coarse-grained
\textit{H}-function is bounded below by zero, $\bar{H}\geq0$; further,
$\bar{H}=0$ if and only if $\bar{P}=\overline{|\Psi|^{2}}$ everywhere
(Valentini 1991a). A decrease of $\bar{H}$ towards its minimum value then
corresponds to an approach of $\bar{P}$ towards $\overline{|\Psi|^{2}}$.

The decrease of $\bar{H}$ corresponds to a `stirring' of the two `fluids' $P$
and $|\Psi|^{2}$ by the same velocity field $\dot{X}$, making $P$ and
$|\Psi|^{2}$ less distinguishable on a coarse-grained level. (This is similar
to the classical Gibbs stirring of two liquids.) As in the corresponding
classical case, the \textit{H}-theorem gives us an insight into the mechanism
whereby equilibrium is approached (for the particular class of initial
conditions specified by (6)). Whether or not equilibrium is actually reached
will depend on the system. It must be assumed that the initial steps towards
equilibrium described by the \textit{H}-theorem are actually completed in
Nature, for appropriately complex systems, so that $\bar{H}(t)\rightarrow0$
and $\bar{P}\rightarrow\overline{|\Psi|^{2}}$.

Given (coarse-grained) equilibrium $\bar{P}=\overline{|\Psi|^{2}}$ for our
ensemble of many-body systems, it is straightforward to show that if a single
particle is extracted from each system and prepared with wavefunction $\psi$,
the resulting ensemble of particle positions has a (coarse-grained)
distribution $\bar{\rho}=\overline{|\psi|^{2}}$ (Valentini 1991a). It is
therefore to be envisaged that the equilibrium distribution seen today for
ensembles of single particles arose via relaxation processes in the complex
systems of which the particles were once part.\footnote{For further details
see Valentini (1991a,b; 1992; 2001a).}

\section{An Estimate for the Relaxation Timescale}

It is important first of all to have a rough, order-of-magnitude estimate for
the timescale over which relaxation to quantum equilibrium takes place. What
we have in mind here is not the time taken to reach equilibrium but the
timescale over which there is a significant \textit{approach} to
equilibrium.\footnote{Cf. the scattering time of classical kinetic theory.}

We may define a relaxation timescale $\tau$ in terms of the rate of decrease
of $\bar{H}(t)$ near $t=0$. Because $\left(  d\bar{H}/dt\right)  _{0}$
vanishes one has to consider $\left(  d^{2}\bar{H}/dt^{2}\right)  _{0}$. Thus
we define $\tau$ by $1/\tau^{2}\equiv-\left(  d^{2}\bar{H}/dt^{2}\right)
_{0}/\bar{H}_{0}$ (Valentini 1992), where (9) gives $\left(  d^{2}\bar
{H}/dt^{2}\right)  _{0}$ in terms of the initial state. Expanding $\dot{X}%
_{0}\cdot\nabla f_{0}$ in a Taylor series within each coarse-graining cell of
volume $\delta V=(\delta x)^{3N}$, it is found that $\left(  d^{2}\bar
{H}/dt^{2}\right)  _{0}=-I(\delta x)^{2}/12+O\left(  (\delta x)^{4}\right)  $
where $I\equiv\int dX\ (|\Psi_{0}|^{2}/f_{0})|\nabla(\dot{X}_{0}\cdot\nabla
f_{0})|^{2}$ (Valentini 2001a). Thus%
\begin{equation}
\tau=\frac{1}{\delta x}\sqrt{\frac{12\bar{H}_{0}}{I}}+O(\delta x)
\end{equation}
For $\delta x$ small compared to the lengthscale over which $\dot{X}_{0}%
\cdot\nabla f_{0}$ varies, $\tau\propto1/\delta x$. Taking $\bar{H}_{0}\sim1$
(a mild disequilibrium) and crudely estimating $I$ one obtains%
\begin{equation}
\tau\sim\frac{1}{\delta x}\frac{m\hslash^{2}}{\left(  \Delta P_{0}\right)
^{3}}%
\end{equation}
where $\Delta P_{0}$ is the (quantum) momentum spread of $\Psi_{0}$ (Valentini
2001a). Given $\tau\propto1/\delta x$ this formula may also be obtained on
dimensional grounds. Note that this result is merely a crude estimate.

\section{Numerical Simulations}

A numerical simulation of relaxation has been performed for the simplest
possible case of an ensemble of independent particles in a one-dimensional box
(Valentini 1992). The results were surprisingly good, given that the particles
cannot move past each other. This simple model is of course an unrealistic
setting for relaxation: one should really consider an ensemble of complicated
systems with many degrees of freedom. Nevertheless the model is instructive.

Consider, then, a one-dimensional box with infinite barriers at $x=0,\ L$ and
energy eigenfunctions $\phi_{n}(x)=\sqrt{2/L}\sin\left(  n\pi x/L\right)  $
($n=1,2,3,.....$) with eigenvalues $E_{n}=\frac{1}{2}(\pi n/L)^{2}$ (units
$m=\hslash=1).$ The box contains an ensemble of independent particles, each
guided by the same wavefunction $\psi(x,t)$. At $t=0$ the wavefunction was
taken to be a superposition of the first $M$ eigenfunctions, with amplitudes
of equal modulus but randomly-chosen phases $\theta_{n}$:%
\[
\psi_{0}(x)=%
{\displaystyle\sum\limits_{n=1}^{M}}
\frac{1}{\sqrt{M}}\phi_{n}(x)\exp(i\theta_{n})
\]
The particles were taken to be uniformly distributed at $t=0$: thus, $\rho
_{0}(x)=1/L$ and of course $\rho_{0}(x)\neq|\psi_{0}(x)|^{2}$. These initial
conditions are sketched in Fig. 1, for the case $M=10$ and $L=100$.

By numerical calculation of the trajectories (given by $\dot{x}=(\partial
/\partial x)\operatorname*{Im}\ln\psi$), the distribution $\rho(x,t)$ at later
times may be determined, while the Schr\"{o}dinger evolution of $\psi(x,t)$ is
just%
\[
\psi(x,t)=%
{\displaystyle\sum\limits_{n=1}^{M}}
\frac{1}{\sqrt{M}}\phi_{n}(x)\exp i(\theta_{n}-E_{n}t)
\]
\newline For $L=100$, $\psi(x,t)$ is periodic in time with period $T_{P}%
=2\pi/E_{1}\approx12,700$ in our units. Because the particles cannot move past
each other, each trajectory must recur with period $T_{P}$ (to ensure that the
equilibrium distribution recurs) and so any $\rho_{0}(x)\neq|\psi_{0}(x)|^{2}$
recurs as well. While this simple model exhibits a strong form of recurrence,
nevertheless for times smaller than the recurrence time the approach to
equilibrium is significant.

An example is shown in Fig. 2, again for ten modes ($M=10$). At $t=120$,
$\rho$ has developed sharp peaks which coincide with the (smooth) maxima of
$|\psi|^{2}$. An experimenter with a `blunt' measuring device, say with
resolution $\delta x=10$, would conclude that $\bar{\rho}$ and $\overline
{|\psi|^{2}}$ are roughly equal. The approach to equilibrium (at $t\ll T_{P}$)
is surprisingly good for such a simple, highly constrained model.\footnote{For
further details see Valentini (1992; 2001a). Note that here the quantum
timescale for the evolution of $|\psi|^{2}$ is $\Delta t\equiv\hslash/\Delta
E\sim70$, where $\Delta E$ is the quantum energy spread.}

The evolution of $\bar{H}(t)$ has also been calculated (with C. Dewdney and B.
Schwentker). An example is shown in Fig. 3, once again for $M=10$, where
$\bar{H}$ is defined with a coarse-graining length $\delta x=1$. The curve
$\bar{H}=\bar{H}(t)$ shows a strict decrease soon after $t=0$ (as it must);
thereafter it decreases steadily but not monotonically, for $t\ll T_{P}$. (Of
course, as the recurrence time is approached, $\bar{H}$ necessarily increases
as the initial conditions are restored.)

It is clear from Fig. 3 that $\bar{H}$ initially decreases on a timescale of
order $\sim100$. This result compares well with our crude estimate (11). For
here $\Delta P_{0}\approx(\pi/\sqrt{3})M/L$ and with $L=100$ and $\delta x=1$
our prediction is $\tau\sim2\times10^{5}/M^{3}$. For $M=10$ we have $\tau
\sim200$ -- in fair agreement with the numerical calculation.

The timescale over which $\bar{H}$ decreases -- which may be quantified in
terms of the time $t_{5\%}$ taken for $\bar{H}$ to decrease by $5\%$ -- has
been calculated (with C. Dewdney and B. Schwentker) for different numbers of
modes, $M=10,\ 15,$ $.....,$ $40$ (with fixed $\delta x=1$), and for different
coarse-graining lengths ranging from $\delta x=0.2$ up to $\delta x=2$ (with
fixed $M=20$). The results agree quite well with our predictions that
$t_{5\%}\propto1/M^{3}$ and $t_{5\%}\propto1/\delta x$ (Dewdney, Schwentker
and Valentini 2001; Valentini 2001a).\footnote{In the first case, upon
averaging over many trials with different random initial phases, a plot of
$\ln t_{5\%}$ versus $\ln M$ was found to have a slope of $-2.6$, compared
with the predicted value of $-3$. In the second case, there are departures
from linearity for $\delta x\gtrsim0.5$ due to the large variation in the
velocity field over some coarse-graining cells, the result $\tau
\propto1/\delta x$ being valid only for small $\delta x$.}

The above model serves an illustrative purpose only. The particles cannot move
past each other, and the approach to equilibrium is limited.
Higher-dimensional models should show a much better approach to equilibrium.

Calculations analogous to the above are currently being performed by Dewdney,
Schwentker and Valentini for a two-dimensional box. Bearing in mind the
generally chaotic nature of the trajectories for this system (Frisk 1997), an
efficient relaxation to equilibrium is to be expected.

\section{Comments on `Typicality'}

A completely different approach to explaining quantum equilibrium is based on
a notion of `typicality' for the initial configuration $X_{0}^{univ}$ of the
whole universe (D\"{u}rr \textit{et al}. 1992a,b). Here, if $\Psi_{0}^{univ}$
is the initial wavefunction of the universe, $|\Psi_{0}^{univ}|^{2}$ is taken
to be the `natural measure' on the space of initial configurations. It is
shown that quantum distributions for measurements on subsystems are obtained
for `almost all' $X_{0}^{univ}$, with respect to the measure $|\Psi_{0}%
^{univ}|^{2}$.

While D\"{u}rr \textit{et al}. give a general proof, their approach may be
illustrated by the case of a universe consisting of an ensemble of $n$
independent subsystems (which could be complicated many-body systems, or
perhaps just single particles), each with wavefunction $\psi_{0}(x)$. Writing
$\Psi_{0}^{univ}=\psi_{0}(x_{1})\psi_{0}(x_{2}).....\psi_{0}(x_{n})$ and
$X_{0}^{univ}=(x_{1},\ x_{2},\ x_{3},\ .....,\ x_{n})$, a choice of
$X_{0}^{univ}$ determines -- for large $n$ -- a distribution $\rho_{0}(x)$
which may or may not equal $|\psi_{0}(x)|^{2}$.

Now it is true that, with respect to the measure $|\Psi_{0}^{univ}|^{2}$, as
$n\rightarrow\infty$ almost all configurations $X_{0}^{univ}$ yield
equilibrium $\rho_{0}=|\psi_{0}|^{2}$ for the subsystems. It might then be
argued that, as $n\rightarrow\infty$, disequilibrium configurations occupy a
vanishingly small volume of configuration space and are therefore
intrinsically unlikely. However, for the above case, with respect to the
measure $|\Psi_{0}^{univ}|^{4}$ almost all configurations $X_{0}^{univ}$
correspond to the \textit{dis}equilibrium distribution $\rho_{0}=|\psi
_{0}|^{4}$. This has led to charges of circularity: that an equilibrium
probability density $|\Psi_{0}^{univ}|^{2}$ is in effect being assumed for
$X_{0}^{univ}$; that the approach amounts to inserting quantum noise into the
initial conditions themselves (Valentini 1996).\footnote{Note that if the word
`typicality' is replaced by `probability', the result of D\"{u}rr \textit{et
al}. (1992a,b) becomes equivalent to the `nesting' property proved by
Valentini (1991a), which states that an equilibrium probability for a
many-body system implies equilibrium probabilities for extracted subsystems,
as briefly discussed above.}

However, there is a certain rationale to D\"{u}rr \textit{et al}.'s approach.
It has been argued (S. Goldstein, private communication) that in statistical
mechanics one always rules out unwanted initial conditions in some way,
usually by deeming them exceptional (or `untypical') with respect to a
specified measure. This is perfectly true. But in the author's opinion, a
class of initial conditions should be selected on \textit{empirical} grounds.
Even today we know only that, for all subsystems probed so far, $\rho
=|\psi|^{2}\pm\epsilon$ where $\epsilon$ is some experimental accuracy. There
is certainly \textit{no} reliable empirical evidence that $\rho=|\psi|^{2}$
exactly -- or even approximately -- in the very early universe, near the big
bang.\footnote{In fact, the horizon problem in cosmology (the uniformity of
the microwave background over regions larger than the classical causal
horizons of a Friedmann universe) suggests that nonequilibrium $\rho\neq
|\psi|^{2}$ may have existed at early times, the resulting nonlocality playing
a role in homogenising the early universe (Valentini 1991b; 1992; 2001a).
Inflation attempts to remove the horizon problem by changing the early
expansion; but as yet there is no satisfactory inflationary theory. Early
quantum disequilibrium offers an alternative approach.}

Note also that pilot-wave theory in exact equilibrium may never be susceptible
to an experimental test of the basic law of motion (the de Broglie guidance
equation); details of the trajectories may well be forever hidden from us,
unless the human brain has an unexpected sensitivity to hidden variables
(Valentini 2001a). Restricting pilot-wave theory to quantum equilibrium is as
artificial as it would be to restrict classical mechanics to thermal
equilibrium; and a huge amount of potentially new physics is lost thereby.

Finally, D\"{u}rr \textit{et al}. focus on initial configurations for the
whole universe while in Sect. 2 we emphasised initial distributions for
subsystems. One might think that the former description is more correct
because in the early universe there are no independent subsystems (that is,
subsystems with their own wavefunction). But even leaving aside the point that
the asymptotic freedom of interactions at high energies allows one to treat
particles at very early times as essentially free, the two modes of
description are really interchangeable: even if there are no independent
subsystems, one can imagine (theoretically) what \textit{would} happen if
(say) single particles were extracted from a gas of interacting particles,
prepared with a certain wavefunction $\psi$, and had their positions measured.
Whether the measured distribution $\rho$ is equal to $|\psi|^{2}$ or not will
depend on the configuration of the whole system prior to the extraction. By
this (purely theoretical) device, one could characterise the total
configuration as `equilibrium' or `nonequilibrium' (Valentini 2001a). Our
Sect. 2 could then be rephrased in terms of configurations rather than
distributions: our `empirical' approach would amount to trying to find out
what sort of initial configuration could explain the statistics observed today
-- and again, while an initial `equilibrium' configuration is a possibility,
there are clearly many `nonequilibrium' configurations at $t=0$ that explain
the observations just as well.

\section{The Early Universe. Suppression of Relaxation at Early Times}

At first sight one might think that, even if the universe did begin in quantum
nonequilibrium, relaxation $\rho\rightarrow|\psi|^{2}$ would take place so
quickly (in the extreme violence of early times) as to erase all record of
nonequilibrium ever having existed. However, closer analysis reveals a
plausible scenario whereby quantum nonequilibrium could survive to the present
day, for some types of particle left over from the big bang, leading to
observable violations of quantum theory.

The scenario has three stages: (i) depending on how the relaxation timescale
$\tau$ scales with temperature $T$, it is possible that relaxation will be
\textit{suppressed} at very early times by the very rapid expansion of space,
in which case (ii) those particles that decouple at very early times will
still be out of quantum equilibrium at the time of decoupling, after which
(iii) the free spreading of particle wavefunctions \textit{stretches} any
residual nonequilibrium up to larger lengthscales that are within reach of
experimental test. This scenario will now be briefly sketched, in this section
and the next.\footnote{For further details see Valentini (2001a,c).}

In the standard hot big bang model, the early universe contains what is
essentially an ideal gas of effectively massless (relativistic) particles. It
is instructive to recall the standard reasoning about how thermal equilibrium
between distinct particle species is achieved: the interaction timescale
$t_{int}$ -- that is, the mean free time between collisions -- must be smaller
than the expansion timescale $t_{exp}\equiv a/\dot{a}\sim(1\ $sec$)(1\ $%
MeV$/kT)^{2}$, where $a(t)\propto t^{1/2}$ is the scale factor and $t$ is the
time since the beginning.\footnote{See any standard cosmology text.} As is
well known, this condition need not always be satisfied: for instance, if the
cross section $\sigma(T)$ falls off faster than $1/T$ at high temperatures
then $t_{int}\sim1/nc\sigma$ -- with a number density $n\sim T^{3}$ (in
natural units where $\hslash=c=1$) -- will fall off slower than $1/T^{2}$ so
that, at sufficiently high temperatures, $t_{int}\gtrsim t_{exp}$ and the
particles do not thermalise. The functions $\sigma=\sigma(T)$ are determined
by particle physics; once they are known, it is possible to determine the
thermal history of the universe, and in particular to deduce when particle
species fall out of thermal contact with each other.

Similar reasoning applies to the relaxation $\rho\rightarrow|\psi|^{2}$. There
will be a temperature-dependent timescale $\tau=\tau(T)$ over which relaxation
takes place. There will also be a \textit{competing} effect due to the
expansion of space: as space expands wavefunctions are stretched, and it is
easy to see in pilot-wave theory that this results in a proportionate
expansion of the disequilibrium lengthscale.

To see this, consider an initial distribution $\rho_{0}(x)$ with
disequilibrium on a small lengthscale $\delta_{0}$. We may write $\rho
_{0}(x)=|\psi_{0}(x)|^{2}f_{0}(x)$, where $\psi_{0}(x)$ has some width
$\Delta_{0}$ and $f_{0}(x)\neq1$ varies over distances $\delta_{0}\ll
\Delta_{0}$ (so that coarse-graining over distances much larger than
$\delta_{0}$ yields an equilibrium distribution). If the wavefunction expands
up to a width $\Delta(t)$, then because $f$ is conserved along trajectories,
and because particles initially separated by a distance $\delta x_{0}$ are
later separated roughly by $\delta x(t)\sim\left(  \Delta(t)/\Delta
_{0}\right)  \delta x_{0}$, it follows that deviations $f\neq1$ occur on an
expanded lengthscale $\delta(t)\sim\left(  \Delta(t)/\Delta_{0}\right)
\delta_{0}$. This is true irrespective of whether the wavefunction expands
because of the expansion of space or simply because of its own natural free
spreading over time.\footnote{This is consistent with the \textit{H}-theorem,
for \textit{H} measures negentropy relative to $|\psi|^{2}$: if $|\psi|^{2}$
itself expands, the absolute disequilibrium lengthscale can increase without
increasing \textit{H}.}

Thus relaxation $\rho\rightarrow|\psi|^{2}$ occurs if $\tau\lesssim t_{exp}$,
while it is \textit{suppressed} if $\tau\gtrsim t_{exp}$. The issue now
depends on how $\tau$ scales with $T$. Unfortunately, there are as yet no
reliable calculations that can tell us this. But we can make a crude estimate.

Given $\tau\propto1/\delta x$, for massless particles $\tau$ may be estimated
on purely dimensional grounds to be $\tau\sim(1/\delta x)\hslash^{2}/c(\Delta
p)^{2}$ where $\Delta p$ is the particle momentum spread. Taking $\Delta p\sim
kT/c$ we have $\tau\sim(1/\delta x)\hslash^{2}c/(kT)^{2}$ which is larger than
$t_{exp}\sim(1\ $sec$)(1\ $MeV$/kT)^{2}$ for all $T$ if $\delta x\lesssim
10l_{P}$ (where $l_{P}$ is the Planck length). However, it would be more
realistic to apply our estimate for $\tau$ to a coarse-graining length $\delta
x$ that is relevant to the energetic processes at temperature $T$, the natural
choice being the thermal de Broglie wavelength $\delta x\sim\hslash c/kT$ (the
typical width of particle wavepackets). We then have $\tau\sim\hslash/kT$ -- a
plausible result, according to which relaxation is suppressed ($\tau\gtrsim
t_{exp}$) if $kT\gtrsim10^{18}\ $GeV$\approx0.1kT_{P}$ or $t\lesssim10t_{P}$
(where $T_{P}$ and $t_{P}$ are the Planck temperature and time). Our simple
estimate suggests that relaxation to quantum equilibrium began one order of
magnitude below the Planck temperature.

Note that in the standard Friedman expansion considered here, the rate
$1/t_{exp}$ at which space is expanding becomes infinite as $t\rightarrow0$.
Of course, our relaxation rate $1/\tau\propto T$ also tends to infinity as
$t\rightarrow0$ (and $T\rightarrow\infty$), but this is offset by the
expansion rate $1/t_{exp}\propto T^{2}$ which tends to infinity even faster.
Note also that while our estimate is very crude and should not be taken too
seriously, it does illustrate the key point that, if $\tau$ decreases with
temperature more slowly than $1/T^{2}$, then relaxation will be suppressed at
very early times.

It is therefore plausible that any particles that decouple soon after the
Planck era will \textit{not} have had time to reach quantum equilibrium, the
extreme violence of that era being offset by the even more extreme expansion
of space. And at the time of decoupling, such particles are expected to show
deviations $\rho\neq|\psi|^{2}$ on a lengthscale $\delta x\sim\hslash
c/kT_{P}=l_{P}\approx10^{-33}\ $cm.

\section{Residual Disequilibrium Today. Experimental Tests}

After decoupling, particle wavefunctions undergo a huge expansion -- partly
due to the expansion of space itself and partly due to the free spreading of
the wavepacket through space. This results in a proportionate stretching of
disequilibrium to much larger lengthscales, as shown
above.\footnote{Decoupling is of course never exact. Because of small residual
interactions at all times, the wavefunctions of relic particles contain tiny
scattering terms that perturb the trajectories -- possibly `re-mixing' the
expanding disequilibrium. Calculation of the effect of scattering by a tenuous
medium shows that re-mixing does \textit{not} happen: the trajectory
perturbations grow only as $t^{1/2}$ and cannot overcome the linear ($\propto
t$) free expansion of the disequilibrium lengthscale (Valentini 2001a,c). We
may therefore safely ignore such residual interactions.}

For example, relic gravitons are believed to decouple at $kT_{dec}\sim
10^{19}\ $GeV or at redshift $z_{dec}\sim T_{dec}/T_{now}\sim10^{32}$. The
subsequent expansion of space alone, by a linear factor of $\sim10^{32}$, will
stretch the disequilibrium lengthscale up to $\sim1\ $mm. Of course, there
seems to be little hope of detecting the $\sim1^{0}\ $K graviton background
directly in the near future -- still less of testing it for violations of
quantum theory. However, it is expected that there are other, more exotic
particles that decoupled soon after $T_{P}$. (Supersymmetry and string theory
predict a plethora of new particles at high energies.) These may be no easier
to detect directly. However, some of them might decay at later times into more
easily detectable particles such as photons, perhaps by annihilation
$X+\bar{X}\rightarrow2\gamma$ or by a supersymmetric decay $X\rightarrow
\gamma+\tilde{\gamma}$ into a photon and a photino. Nonequilibrium for the
parent particles should yield nonequilibrium for the decay products as well.
Thus, we would suggest testing the Born rule $\rho=|\psi|^{2}$ for photons
produced by the decay of exotic relic particles from the Planck era. For
example, in a two-slit interference experiment, such photons might produce an
interference pattern that deviates from the quantum prediction.

Experiments are under way searching for exotic relic particles supposed to
make up the `dark matter' pervading the universe. Some of these experiments
involve searching for decay photons. However, the usual dark matter candidates
(such as neutrinos, neutralinos, axions or gravitinos) are expected to
decouple much later than $t_{P}$. Nevertheless, our knowledge of particle
physics beyond the standard model is so uncertain that, for all we know, there
might exist an appropriate relic particle $X$ -- that decouples soon after
$t_{P}$ and partially accounts for dark matter. If such particles were
detected, they or their decay products would be our prime candidates for
particles violating the Born rule.

In the meantime one might also consider the relic photons that make up the
microwave background. But why consider particles that decoupled at
$t\sim10^{5}\ $yr -- \textit{much} later than $t_{P}$ -- when our estimates
suggest that any disequilibrium will have been erased? One answer is that the
hugely expanded wavepackets of relic particles provide an interesting test of
the Born rule in extreme conditions. On general grounds (irrespective of the
suggestions made here about hidden variables and early nonequilibrium), it
would be worth testing quantum theory in such unusual circumstances, where
quantum probabilities have spread over intergalactic distances. Relic photons
decouple at $kT_{dec}\sim1\ $eV ($z_{dec}\sim10^{3}$) and at decoupling their
wavepackets have widths of order $\sim\hslash c/(1\ $eV)$\sim10^{-5}\ $cm. If
the packets have (as is widely assumed) spread essentially freely at the speed
of light for $\sim10^{10}$ years then today they will have a width
$\sim10^{28}\ $cm and so the packets will have expanded by $\sim10^{33}$
(ignoring the relatively small effect of the expansion of space by a factor
$\sim10^{3}$). It can be argued -- on general grounds, independent of
pilot-wave theory -- that for cosmological microwave photons the quantum
probability today on lengthscales $\sim1\ $cm could contain traces of
corrections to the Born rule which may have existed (for whatever reason) on
the Planck scale at the time of decoupling (Valentini 2001a,c).\footnote{Our
point here is that the line of argument given in this paper has led us to
propose experimental tests that are actually worthwhile in their own right,
because they would probe quantum theory in new and extreme conditions.}

As suggested elsewhere (Valentini 1996), it would then be worthwhile to test
quantum theory for photons from the microwave background. In a two-slit
interference experiment with single relic photons, the usual quantum
interference pattern -- given by $\rho=|\psi|^{2}$ -- might show an
\textit{anomalous blurring}.\footnote{Incoming photons will have very nearly
plane wavepackets, so there is no ambiguity as to what quantum theory would
predict in such an experiment. Note also that it has recently become possible
to detect single photons in the far-infrared region (Komiyama \textit{et al}. 2000).}

Generally, corrections to the Born rule for any photons from deep space would
produce a number of observable anomalies. In particular, the functioning of
some astronomical instruments might be affected: diffraction-grating
spectrometers could produce unreliable readings, and the diffraction-limited
performance of some telescopes might be impaired.\footnote{For details see
Valentini (2001a).}

\section{Outlook}

It is clear that much remains to be done to develop the above ideas fully.
Other approaches to subquantum statistical mechanics (based for example on
external perturbations) remain to be developed. Properties of the trajectories
such as ergodicity and mixing (in a rigorous sense) should be investigated.

The process of relaxation to quantum equilibrium should also be studied
further, in particular in the early universe. It remains to be seen if the
relaxation timescale $\tau$ really does decrease with temperature more slowly
than $1/T^{2}$, so that relaxation is suppressed at very early times. An
experimental test of the Born rule $\rho=|\psi|^{2}$ for relic cosmological
particles seems feasible, in particular for photons from the microwave
background (Valentini 2001a,c).

As for the issue of chance in physics, the central conclusion of this work is
that, like the cosmic microwave background, \textit{quantum noise is a remnant
of the big bang}. And just as the microwave background has been found to have
small nonuniformities in temperature, so the `quantum background' -- the
quantum noise that pervades our universe -- may have small deviations from the
Born rule $\rho=|\psi|^{2}$, and should be probed experimentally.

\textbf{Acknowledgements.} I am grateful to audiences at the Universities of
Portsmouth and Utrecht for their comments, to Janneke van Lith for a critical
reading of section 2, and to Chris Dewdney for preparing the figures.

\begin{center}
\textbf{FIGURE CAPTIONS}
\end{center}

\textbf{Fig. 1.} Initial conditions at $t=0$. We have plotted $|\psi|^{2}$ for
a superposition of ten modes with amplitudes of equal modulus and random
phase. The initial density $\rho$ is taken to be uniform.

\textbf{Fig. 2.} At $t=120$. There is a strong coincidence between the sharp
peaks of $\rho$ and the smooth peaks of $|\psi|^{2}$, indicating an
approximate approach to equilibrium on a coarse-grained level.

\textbf{Fig. 3.} Plot of the coarse-grained $\bar{H}$-function against time.

\begin{center}
\textbf{REFERENCES}
\end{center}

G. Bacciagaluppi and A. Valentini (2001), in preparation.

D. Bohm (1952a), A suggested interpretation of the quantum theory in terms of
`hidden' variables. I, \textit{Physical Review} 85, 166--179.

D. Bohm (1952b), A suggested interpretation of the quantum theory in terms of
`hidden' variables. II, \textit{Physical Review} 85, 180--193.

L. de Broglie (1928), La nouvelle dynamique des quanta, in: \textit{Electrons
et Photons}, Gauthier-Villars, Paris.

C. Dewdney, B. Schwentker and A. Valentini (2001), in preparation.

D. D\"{u}rr, S. Goldstein and N. Zangh\`{i} (1992a), Quantum equilibrium and
the origin of absolute uncertainty, \textit{Journal of Statistical Physics}
67, 843--907.

D. D\"{u}rr, S. Goldstein and N. Zangh\`{i} (1992b), Quantum mechanics,
randomness, and deterministic reality, \textit{Physics Letters} A 172, 6--12.

H. Frisk (1997), Properties of the trajectories in Bohmian mechanics,
\textit{Physics Letters} A 227, 139--142.

S. Komiyama, O. Astafiev, V. Antonov, T. Kutsuwa and H. Hirai (2000), A
single-photon detector in the far-infrared range, \textit{Nature} 403, 405--407.

R. Tolman (1938), \textit{The Principles of Statistical Mechanics}, Oxford.

A. Valentini (1991a), Signal-locality, uncertainty, and the subquantum
\textit{H}-theorem. I, \textit{Physics Letters} A 156, 5--11.

A. Valentini (1991b), Signal-locality, uncertainty, and the subquantum
\textit{H}-theorem. II, \textit{Physics Letters} A 158, 1--8.

A. Valentini (1992), On the pilot-wave theory of classical, quantum and
subquantum physics, PhD thesis, SISSA/ISAS, Trieste, Italy.

A. Valentini (1996), Pilot-wave theory of fields, gravitation and cosmology,
in: \textit{Bohmian Mechanics and Quantum Theory: an Appraisal}, 45--66 (eds.
J.T. Cushing \textit{et al.}), Kluwer.

A. Valentini (1997), On Galilean and Lorentz invariance in pilot-wave
dynamics, \textit{Physics Letters} A 228, 215--222.

A. Valentini (2001a), \textit{Pilot-Wave Theory: an Alternative Approach to
Modern Physics}, Springer, to be published.

A. Valentini (2001b), to be submitted.

A. Valentini (2001c), to be submitted.
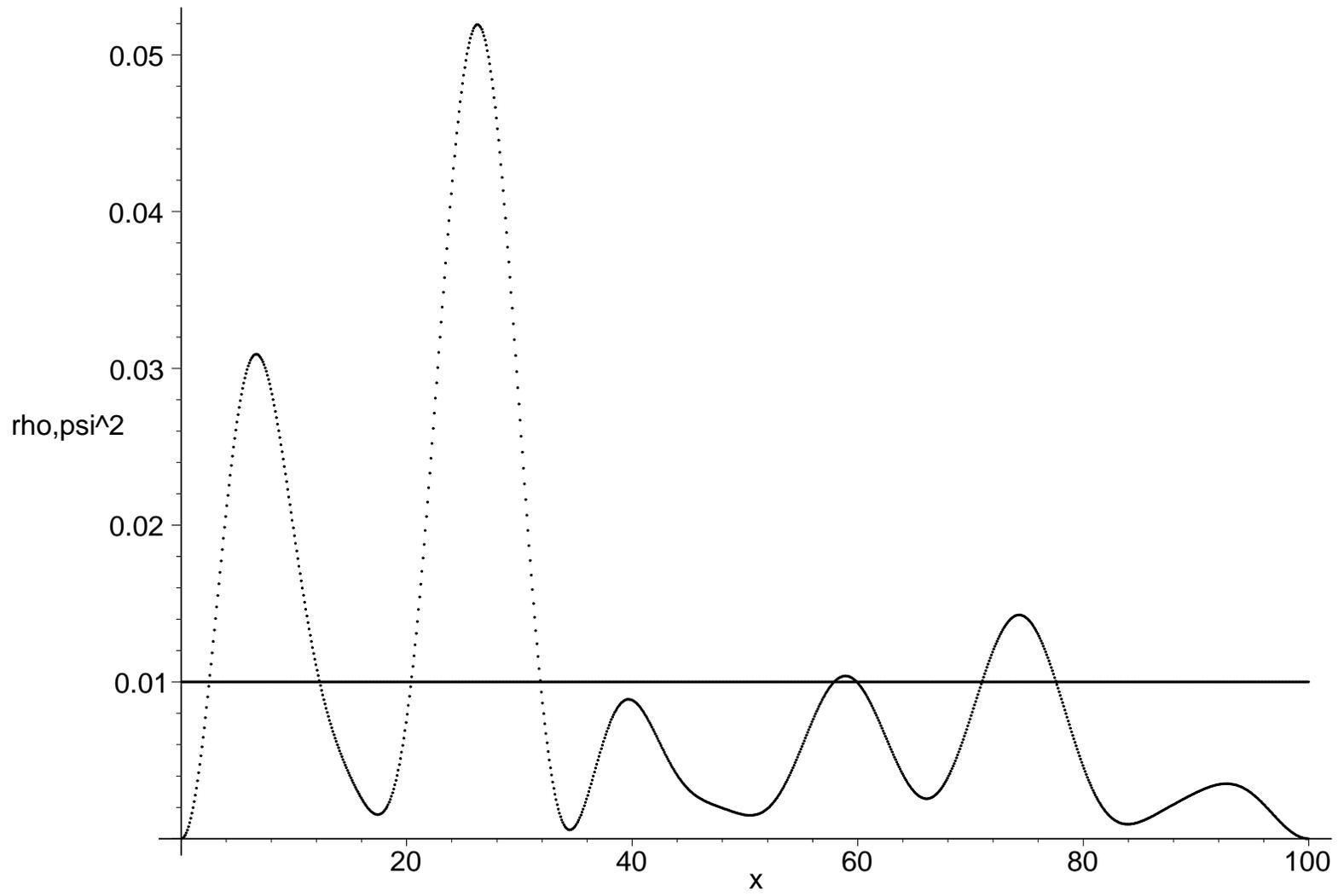
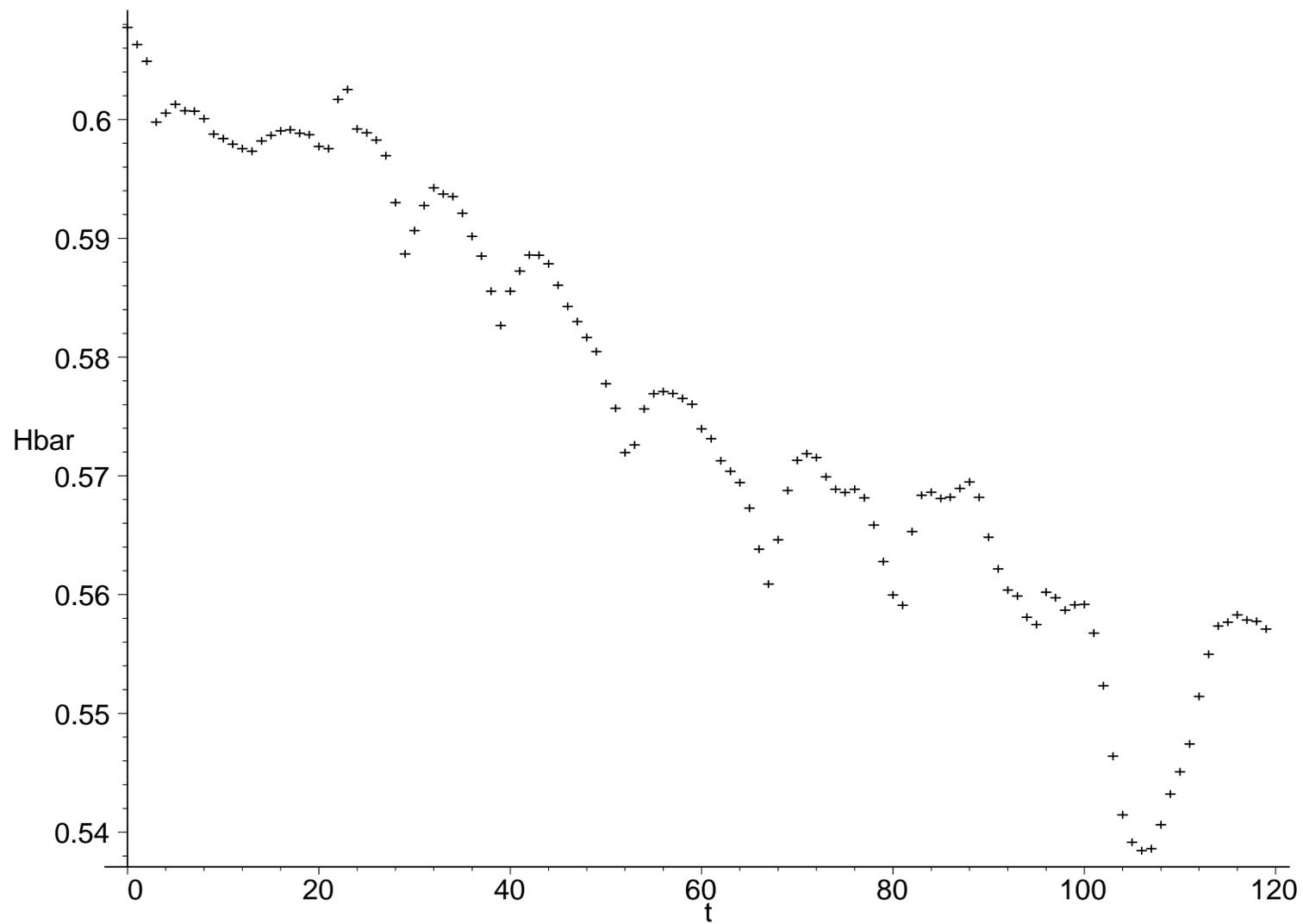
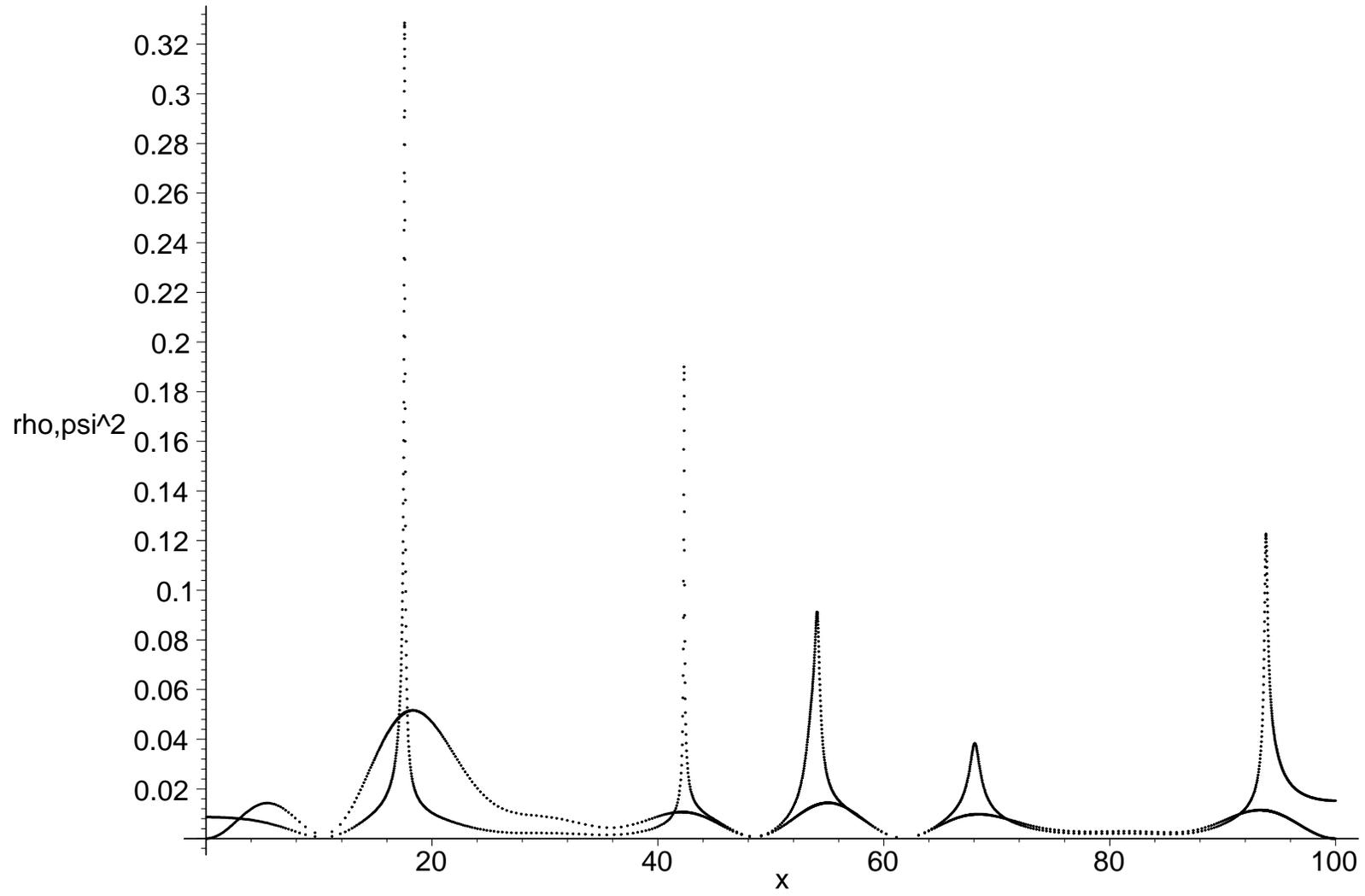
\end{document}